# A Note on the Convergence of the Gaussian Mean Shift Algorithm

Hien D. Nguyen[1]

**Abstract**

Mean shift (MS) algorithms are popular methods for mode finding in pattern analysis. Each MS algorithm can be phrased as a fixed-point iteration scheme, which operates on a kernel density estimate (KDE) based on some data. The ability of an MS algorithm to obtain the modes of its KDE depends on whether or not the fixed-point scheme converges. The convergence of MS algorithms have recently been proved under some general conditions via first principle arguments. We complement the recent proofs by demonstrating that the MS algorithm operating on a Gaussian KDE can be viewed as an MM (minorization–maximization) algorithm, and thus permits the application of convergence techniques for such constructions. For the Gaussian case, we extend upon the previously results by showing that the fixed-points of the MS algorithm are all stationary points of the KDE in cases where the stationary points may not necessarily be isolated.

## 1. Introduction

Ever since their introduction by [11], and then extension by [7], mean shift (MS) algorithms have become mainstay tools in the kits of pattern analysts. There have been numerous articles on their applications to various problem

[1]Department of Mathematics and Statistics, La Trobe University, Melbourne, Australia

domains. A recent and thorough review on MS algorithms and their variants can be found in [6].

Following the exposition of [2], we can describe the MS algorithms as follows. Let $\boldsymbol{x} \in \mathbb{R}^d$ ($d \in \mathbb{N}$) and let $K$ be a kernel on $\boldsymbol{x}$, satisfying the properties: $K(\boldsymbol{x}) \geq 0$ for all $\boldsymbol{x}$ and $\int_{\mathbb{R}^d} K(\boldsymbol{x}) \, \mathrm{d}\boldsymbol{x} = 1$ ($K$ is a probability density function on $\mathbb{R}^d$), $\lim_{\|\boldsymbol{x}\| \to \infty} \|\boldsymbol{x}\|^d K(\boldsymbol{x}) = 0$ ($K$ vanishes), $\int_{\mathbb{R}^d} \boldsymbol{x} K(\boldsymbol{x}) \, \mathrm{d}\boldsymbol{x} = \boldsymbol{0}$ ($K$ has zero expectation), and $\int_{\mathbb{R}^d} \boldsymbol{x}\boldsymbol{x}^\top K(\boldsymbol{x}) \, \mathrm{d}\boldsymbol{x} = c_K \mathbf{I}$ ($K$ has spherical covariance), where $\boldsymbol{0}$ is the zero vector and $\mathbf{I}$ is the identity matrix. Here $\|\cdot\|$ is the Euclidean norm, $c_K$ is a constant that depends only on $K$, and $\top$ indicates matrix transposition.

Let $\mathbb{X} = \{\boldsymbol{x}_i\}_{i=1}^n$ be a ssample from some probability space, such that $\boldsymbol{x}_i \in \mathbb{R}^d$ for each $i \in [d]$ ($[d] = \{1, ..., d\}$). We say that $\hat{f}$ is the kernel density estimator (KDE) constructed from the sample $\mathbb{X}$ if it has the form

$$\hat{f}(\boldsymbol{x}) = \frac{1}{n} \sum_{i=1}^n K_{\mathbf{H}}(\boldsymbol{x} - \boldsymbol{x}_i), \tag{1}$$

where $K_{\mathbf{H}}(\boldsymbol{x}) = |\mathbf{H}|^{-1/2} K(\mathbf{H}^{-1/2}\boldsymbol{x})$ and $\mathbf{H}$ is a symmetric and positive definite matrix which we call the bandwidth matrix.

Let $k$ be a univariate non-negative, non-increasing, and piecewise continuous function that satisfies the condition $\int_0^\infty k(x) \, \mathrm{d}x < \infty$, and the relationship $K(\boldsymbol{x}) = c_{kd} k\left(\|\boldsymbol{x}\|^2\right)$. Here $c_{kd}$ is a normalizing constant that satisfies $\int_{\mathbb{R}^d} c_{kd} k\left(\|\boldsymbol{x}\|^2\right) \mathrm{d}\boldsymbol{x} = 1$. We refer to $k$ as the profile of the kernel $K$. A common and popular kernel choice is to take $K$ to be Gaussian, with the form

$$K(\boldsymbol{x}) = \frac{\exp\left(-\|\boldsymbol{x}\|^2/2\right)}{(2\pi)^{d/2}},$$

which we can write in profile using $c_{kd} = 1/(2\pi)^{d/2}$ and $k(x) = \exp(-x/2)$. Suppose that we make a further simplification by taking the bandwidth to be spherical with the form $\mathbf{H} = h^2 \mathbf{I}$, where $h > 0$. We can then write (1) in the

alternative form

$$\hat{f}(\boldsymbol{x}) = \frac{c_{kd}}{nh^d} \sum_{i=1}^{n} k\left(\left\|\frac{\boldsymbol{x}-\boldsymbol{x}_i}{h}\right\|^2\right). \tag{2}$$

If we assume that $k$ is also differentiable (which the Gaussian profile is), then [8] showed that one can write the gradient of (2) as

$$\nabla \hat{f}(\boldsymbol{x}) = \frac{2c_{kd}}{nh^{d+2}} \left[\sum_{i=1}^{n} g\left(\left\|\frac{\boldsymbol{x}-\boldsymbol{x}_i}{h}\right\|^2\right)\right] \left[\frac{\sum_{i=1}^{n} \boldsymbol{x}_i g\left(\left\|\frac{\boldsymbol{x}-\boldsymbol{x}_i}{h}\right\|^2\right)}{\sum_{i=1}^{n} g\left(\left\|\frac{\boldsymbol{x}-\boldsymbol{x}_i}{h}\right\|^2\right)} - \boldsymbol{x}\right], \tag{3}$$

where $g(x) = -\mathrm{d}k/\mathrm{d}x$ is the profile of some kernel $G$ ($K$ is said to be the shadow of $G$) and $\nabla$ is the gradient operator. That is $G(\boldsymbol{x}) = c_{gd} g\left(\|\boldsymbol{x}\|^2\right)$ is a kernel with normalizing constant $c_{gd}$. The second bracket on the right-hand side of (3) is referred to as the MS vector $\boldsymbol{\mu}(\boldsymbol{x})$ and we can see—via the usual first-order conditions (FOCs) of multivariate calculus—that the problem of obtaining the modes of the KDE $\hat{f}$ can be translated to finding the appropriate roots of the gradient equation $\nabla \hat{f} = \boldsymbol{0}$, or more simply $\boldsymbol{\mu}(\boldsymbol{x}) = \boldsymbol{0}$. Assuming that the equation

$$\boldsymbol{\mu}(\boldsymbol{x}) = \frac{\sum_{i=1}^{n} \boldsymbol{x}_i g\left(\left\|\frac{\boldsymbol{x}-\boldsymbol{x}_i}{h}\right\|^2\right)}{\sum_{i=1}^{n} g\left(\left\|\frac{\boldsymbol{x}-\boldsymbol{x}_i}{h}\right\|^2\right)} - \boldsymbol{x}$$

has a fixed point, the MS algorithm for the KDE $\hat{f}$ that is initialized at some value $\boldsymbol{m}^{(0)}$ can be defined as follows. Let $\boldsymbol{m}^{(r)}$ be the $r$th iterate of the algorithm; at the $(r+1)$th iteration, perform the update:

$$\boldsymbol{m}^{(r+1)} = \boldsymbol{\mu}\left(\boldsymbol{m}^{(r)}\right) + \boldsymbol{m}^{(r)} = \frac{\sum_{i=1}^{n} \boldsymbol{x}_i g\left(\left\|\frac{\boldsymbol{m}^{(r)}-\boldsymbol{x}_i}{h}\right\|^2\right)}{\sum_{i=1}^{n} g\left(\left\|\frac{\boldsymbol{m}^{(r)}-\boldsymbol{x}_i}{h}\right\|^2\right)}. \tag{4}$$

The algorithm is terminated once the condition $\left\|\boldsymbol{m}^{(r+1)} - \boldsymbol{m}^{(r)}\right\| < \text{TOL}$ is reached, for some small $\text{TOL} > 0$. We generally initialize $\boldsymbol{m}^{(0)}$ at each of the

sample elements of $\mathbb{X}$ and perform $n$ MS algorithm runs in parallel.

A problem that arises in the study of MS algorithms is the determination of whether and when the iterative scheme (4) is convergent. The recent works of [1] and [2] have made great progress in this direction of research. In particular, the conditions for convergence of the MS algorithm on a Gaussian KDE to isolated stationary points is established in [2].

The insufficiencies of previous proof techniques of [8] and [15] are discussed in [2]. Further, the insufficiencies of techniques relying on comparisons of the MS algorithm to the EM (expectation–maximization; [9]) of [5] and [14] are dismissed due to the perceived lack in regularity fulfillment of the conditions from [21] leading to the potential pathologies that are discussed in [4]. In this article, we demonstrate that the EM algorithm approach should not be dismissed and can indeed lead to proofs of convergence in the MS sequence to stationary points, both in the isolated and unknown cases.

We approach the problem via the MM (minorization–maximization; [12]) algorithm framework, which may be viewed as a generalization of the EM algorithms. Recent reviews of the MM algorithm literature can be found in [17] and [19]. We demonstrate the fact that in the Gaussian KDE case, the iterative scheme (4) constitutes an MM algorithm. Convergence of MM algorithms under various conditions have been proved in [20], [18], and [13, Ch. 12]. Demonstrating that there exists an MM algorithm corresponding to the MS algorithm satisfying the regularity conditions of [18], and [13, Ch. 12], we can conclude that the MS algorithm is convergent in both cases where the stationary points are isolated or otherwise.

The article proceeds as follows. The MM algorithm framework is introduced and the conditions for convergence of MM algorithms are presented in Section 2. In Section 3, an MM algorithm with the same iterates as the MS algorithm for the Gaussian KDE is derived and the main result of the article is presented. Conclusions are drawn in Section 4.

## 2. The MM Algorithm Framework

Let $O(\boldsymbol{\theta})$ be some objective function that one wishes to maximize, where $\boldsymbol{\theta} \in \Theta \subset \mathbb{R}^q$ for some $q \in \mathbb{N}$. Suppose that $O$ is difficult to operate on for some reason (e.g. when the solution to the FOCs cannot be obtained in closed form). Suppose that $M(\boldsymbol{\theta}; \boldsymbol{v})$ is a surrogate function in $\boldsymbol{\theta}$ that exists for every $\boldsymbol{v} \in \Theta$, which is easier to operate on than $O$. If $M$ satisfies the conditions (i) $M(\boldsymbol{v}; \boldsymbol{v}) = O(\boldsymbol{v})$ and (ii) $M(\boldsymbol{\theta}; \boldsymbol{v}) \leq O(\boldsymbol{\theta})$ for every $\boldsymbol{\theta}, \boldsymbol{v} \in \Theta$, then we say that $M$ is a minorizer of $O$ at $\boldsymbol{v}$ (or that $M$ minorizes $O$ at $\boldsymbol{v}$). Define $\boldsymbol{\theta}^{(0)}$ to be some initial value and $\boldsymbol{\theta}^{(r)}$ to be the $r$th iterate of the MM algorithm. We can define the MM algorithm for maximizing $O$ via the iteration scheme

$$\boldsymbol{\theta}^{(r+1)} = \arg\max_{\boldsymbol{\theta} \in \Theta} M\left(\boldsymbol{\theta}; \boldsymbol{\theta}^{(r)}\right). \qquad (5)$$

Together, the definitions of the minorizer and the iterate scheme (5) produce the chain of inequalities

$$O\left(\boldsymbol{\theta}^{(r)}\right) = M\left(\boldsymbol{\theta}^{(r)}; \boldsymbol{\theta}^{(r)}\right) \leq M\left(\boldsymbol{\theta}^{(r+1)}; \boldsymbol{\theta}^{(r)}\right) \leq O\left(\boldsymbol{\theta}^{(r+1)}\right),$$

implying that the sequence of functional evaluates $O\left(\boldsymbol{\theta}^{(r)}\right)$ is monotonically increasing in $r$.

Define the directional derivative of $O$ at $\boldsymbol{\theta}$ in the direction $\boldsymbol{\delta}$ as

$$O'(\boldsymbol{\theta}; \boldsymbol{\delta}) = \lim_{\lambda \downarrow 0} \frac{O(\boldsymbol{\theta} + \lambda \boldsymbol{\delta}) - O(\boldsymbol{\theta})}{\lambda},$$

and define a stationary point of $O$ to be any point $\boldsymbol{\theta}^*$ that satisfies the condition

$O'\left(\boldsymbol{\theta}^{*};\boldsymbol{\delta}\right) \geq 0$, for all $\boldsymbol{\delta}$ such that $\boldsymbol{\theta}+\boldsymbol{\delta} \in \Theta$. Starting from some initial value $\boldsymbol{\theta}^{(0)}$, we can define $\boldsymbol{\theta}^{(\infty)} = \lim_{r\to\infty}\boldsymbol{\theta}^{(r)}$ to be a limit point of the MM algorithm (5). Via Theorem 1 of [18], we have the following result regarding limit points.

**Theorem 1** (Razaviyayn et al., 2013, Thm. 1). *If $O$ is differentiable, then every limit point $\boldsymbol{\theta}^{(\infty)}$ of the iterative scheme (5) is a stationary point of $O$.*

Note that Theorem 1 states that if a limit point is obtainable from some initial value $\boldsymbol{\theta}^{(0)}$, then it is a stationary point of the objective function. The theorem does not provide a result regarding the existence of a limit point for any initial value $\boldsymbol{\theta}^{(0)}$. We can obtained such a result by making some stronger assumptions on $O$.

Say that $O$ is coercive if $\lim_{\|\boldsymbol{\theta}\|\to\infty} O\left(\boldsymbol{\theta}\right) = -\infty$ and that a point $\boldsymbol{v}$ in a set $\Upsilon$ is isolated if and only if there exists a $\rho > 0$ such that $\Upsilon \cap B\left(\boldsymbol{v},\rho\right) = \{\boldsymbol{v}\}$, where $B\left(\boldsymbol{v},\rho\right)$ is a ball around $\boldsymbol{v}$ of radius $\rho$. Proposition 12.4.4 of [13] provides the following result.

**Theorem 2** (Lange, 2013, Prop. 12.4.4). *Suppose that the set of stationary points of $O$ are isolated and that $O$ is differentiable and coercive. If the minorizer $M\left(\cdot;\boldsymbol{v}\right)$ is strictly concave for every $\boldsymbol{v} \in \Theta$, then any sequence of MM iterates $\boldsymbol{\theta}^{(r)}$ (starting from any initial value $\boldsymbol{\theta}^{(0)}$) generated via scheme (5) possesses a limit point $\boldsymbol{\theta}^{(\infty)}$, and $\boldsymbol{\theta}^{(\infty)}$ is a stationary point of $O$.*

### 3. Gaussian Mean Shift as an MM Algorithm

In order to apply Theorems 1 and 2, we require the representation of the iteration scheme (4)—for a Gaussian KDE—as an MM scheme of form (5), for some objective function $O$ and minorizer $M$. We first start by considering that the derivative of the Gaussian profile $k$ has the form $\mathrm{d}k/\mathrm{d}x = -\exp\left(-x/2\right)/2$ and thus $g\left(x\right) = -\mathrm{d}k/\mathrm{d}x \propto k\left(x\right)$. Hence, we have the fact that if $k$ is the profile of a Gaussian kernel then $g$ is also the profile of a Gaussian kernel. We thus require the derivation of an MM algorithm that has iterates of the form

$$\boldsymbol{m}^{(k+1)} = \frac{\sum_{i=1}^{n} \boldsymbol{x}_i \exp\left(-\frac{1}{2}\left\|\frac{\boldsymbol{m}^{(r)}-\boldsymbol{x}_i}{h}\right\|^2\right)}{\sum_{i=1}^{n} \exp\left(-\frac{1}{2}\left\|\frac{\boldsymbol{m}^{(r)}-\boldsymbol{x}_i}{h}\right\|^2\right)}.$$

Let

$$\hat{f}(\boldsymbol{m}) = \frac{\left(2\pi h^2\right)^{-d/2}}{n} \sum_{i=1}^{n} \exp\left(-\frac{1}{2}\left\|\frac{\boldsymbol{m}-\boldsymbol{x}_i}{h}\right\|^2\right) \tag{6}$$

be a Gaussian KDE, or alternatively, the probability density function (with respect to $\boldsymbol{m}$) of a homoskedastic $n$-component Gaussian mixture model, where each component has covariance $h^2 \mathbf{I}$, prior probabilities $1/n$, and means at the $n$ values $\boldsymbol{x}_i$ in the sample $\mathbb{X}$. See [16, Ch. 3] regarding Gaussian mixture models. Suppose that we wish to obtain the local maxima of $\hat{f}$. Since the logarithm is strictly increasing, the problem of obtaining the local maxima of $\hat{f}$ is equivalent to the problem of obtaining the local maxima of

$$l(\boldsymbol{m}) = \log \sum_{i=1}^{n} \exp\left(-\frac{1}{2}\left\|\frac{\boldsymbol{m}-\boldsymbol{x}_i}{h}\right\|^2\right) - \log n - \frac{d}{2}\log\left(2\pi h^2\right). \tag{7}$$

We note that all of the components of $l$ are smooth in $\boldsymbol{m}$ and thus one can obtain the stationary points (including the local maxima, minima, and saddle points) of $l$ via the FOC $\nabla l = \mathbf{0}$. Unfortunately, one cannot obtain such solutions in closed form, thus an alternative approach is required. Such an alternative is via an iterative MM algorithm.

Observe that the difficulty in operating with (7) is the log-sum-exp form in the first expression (cf. [3, Sec. 3.1]). The following minorizer from [22] is useful for operating with such expressions.

**Lemma 1.** *The objective function* $O(\boldsymbol{\theta}) = \log\left(\sum_{i=1}^{q} \theta_i\right)$ *is minorized by*

$$M(\boldsymbol{\theta}; \boldsymbol{v}) = \sum_{i=1}^{q} \frac{v_i}{\sum_{j=1}^{q} v_j} \log\left(\frac{\sum_{j=1}^{q} v_j}{v_i} \theta_i\right),$$

*for* $\boldsymbol{\theta}, \boldsymbol{v} \in [0, \infty)^q$.

Let $\boldsymbol{m}^{(r)}$ be the $r$th iterate of the MM algorithm. We obtain the minorizer

$$M\left(\boldsymbol{m}; \boldsymbol{m}^{(r)}\right) = C - \frac{1}{2h^2} \sum_{i=1}^{n} \frac{\exp\left(-\frac{1}{2}\left\|\frac{\boldsymbol{m}^{(r)} - \boldsymbol{x}_i}{h}\right\|^2\right)}{\sum_{j=1}^{n} \exp\left(-\frac{1}{2}\left\|\frac{\boldsymbol{m}^{(r)} - \boldsymbol{x}_i}{h}\right\|^2\right)} \|\boldsymbol{m} - \boldsymbol{x}_i\|^2 \quad (8)$$

of (7) in terms of $\boldsymbol{m}$, by setting

$$\theta_i = \exp\left(-\frac{1}{2}\left\|\frac{\boldsymbol{m} - \boldsymbol{x}_i}{h}\right\|^2\right) \text{ and } v_i = \exp\left(-\frac{1}{2}\left\|\frac{\boldsymbol{m}^{(r)} - \boldsymbol{x}_i}{h}\right\|^2\right),$$

for $i \in [n]$, and applying Lemma 1 to (7). Here,

$$C = -\log n - \frac{d}{2}\log\left(2\pi h^2\right) - \sum_{i=1}^{n} \frac{v_i}{\sum_{j=1}^{n} v_j} \log\left(\frac{v_i}{\sum_{j=1}^{n} v_j}\right)$$

is a constant that does not depend on $\boldsymbol{m}$.

For fixed $\boldsymbol{m}^{(r)}$, (8) is a negative quadratic in $\boldsymbol{m}$ and is thus strictly concave. The gradient of (8) has the form

$$\nabla M = -\frac{1}{h^2} \sum_{i=1}^{n} \frac{\exp\left(-\frac{1}{2}\left\|\frac{\boldsymbol{m}^{(r)} - \boldsymbol{x}_i}{h}\right\|^2\right)}{\sum_{j=1}^{n} \exp\left(-\frac{1}{2}\left\|\frac{\boldsymbol{m}^{(r)} - \boldsymbol{x}_i}{h}\right\|^2\right)} (\boldsymbol{m} - \boldsymbol{x}_i)$$

which yields the unique FOC solution

$$\boldsymbol{m}^* = \frac{\sum_{i=1}^n \boldsymbol{x}_i \exp\left(-\frac{1}{2}\left\|\frac{\boldsymbol{m}^{(r)}-\boldsymbol{x}_i}{h}\right\|^2\right)}{\sum_{i=1}^n \exp\left(-\frac{1}{2}\left\|\frac{\boldsymbol{m}^{(r)}-\boldsymbol{x}_i}{h}\right\|^2\right)}.$$

We can then construct an MM algorithm for maximizing $l$ of form (4) by repeating the iterations $\boldsymbol{m}^{(r+1)} = \boldsymbol{m}^*$. Finally, it is elementary to check that $l$ is coercive in $\boldsymbol{m}$. Thus, we have the conditions of Theorems 1 and 2 fulfilled. Our main result then follows.

**Theorem 3.** *If $g$ is Gaussian (i.e. $g(x) \propto \exp(-x/2)$), then (i) the fixed-points of the MS algorithm iterates $\boldsymbol{m}^{(r)}$ (as defined by (4)) are stationary points of the Gaussian mixture log-density (7); (ii) if the stationary points of (7) are also isolated, then every sequence of MS algorithm iterates $\boldsymbol{m}^{(r)}$ possesses a limit point which is also a stationary point of (7); and (iii) each sequence of Gaussian mixture log-density evaluates $l\left(\boldsymbol{m}^{(r)}\right)$ is monotonically increasing in $r$.*

We note that the MS algorithm for seeking the roots of the Gaussian KDE $\hat{f}$ can in fact be represented as an MM algorithm that monotonically seeks increases in the Gaussian mixture log-density $l$ (cf. Theorem 3, Part (iii)). This is a good result as it implies that the algorithm naturally avoids attraction of iterates towards local minima of $l$, which are contradictory to the purpose of the MS algorithm.

## 4. Conclusions

In the main result of [2], it was established that the sequences of modal estimates that are obtained via MS algorithms are convergent for general KDEs. The author establish the result under an assumption that the stationary points of the KDE in use are isolated. Two lemmas (9 and 10) are provided as mechanisms for checking the assumption of isolated stationary points for Gaussian KDEs.

The results of [2] were obtained via careful linear algebraic manipulations and first-principle arguments regarding the convergence of sequences. The main result of our article offers an alternative and complementary perspective on convergence of the MS algorithms—to that of [2]—for the Gaussian case. It is also fully compatible with the stationary point isolation lemmas for Gaussian KDEs.

In our approach, we demonstrated that the Gaussian MS algorithm is an instance of an MM algorithm and thus allows for the establishment of convergence via the available results for such constructions. Along with the comparable result to that of [2] regarding the case when the stationary points are isolated (i.e. Theorem 3, Part (ii)), we also present a convergence result for situations when isolation of stationary points cannot be established (i.e. Theorem 3, Part (i)).

We note that the representation of MS algorithms as optimization schemes has a history. For example, [5] and [14] considered MS algorithms as EM-type algorithms and [10] considered MS algorithms as bound optimization schemes. We hope that the insights from this article can be further applied to establish the convergence of other MS algorithms and variants in the future.